\documentclass{IEEE_lsens}
\usepackage{textcomp}

\usepackage[noadjust]{cite}

\ifCLASSINFOpdf
\else
\fi
\usepackage[T1]{fontenc} 
\usepackage{amsmath}
\usepackage{graphicx}
\graphicspath{ {./Figures/} }
\usepackage{xcolor}
\usepackage[cmintegrals]{newtxmath}
\usepackage{bm}

\usepackage{array}
\usepackage{url}
\ifCLASSINFOpdf
\else
\fi
\providecommand{\hypersetup}[1]{\relax}

\hypersetup{pdftitle={Low cost pH sensor fabricated on scotch tape},
pdfsubject={Typesetting},
pdfauthor={Siddharth Tallur},
pdfkeywords={Class, IEEE, IEEE\_lsens, IEEE Sensors Letters, LaTeX, Typesetting, TeX}}
\begin{document}

\markboth{Vol.~1, No.~3, July~2017}{0000000}

\IEEELSENSarticlesubject{Sensor materials and solid-state sensors}

%
\title{Low cost passive pH sensor fabricated on scotch tape}

%
\author{\IEEEauthorblockN{Ayan~Majumder\IEEEauthorrefmark{1}, Subrat~Kumar~Pradhan\IEEEauthorrefmark{1}, Kasturi~Saha\IEEEauthorrefmark{1},
and~Siddharth~Tallur\IEEEauthorrefmark{1}}
\IEEEauthorblockA{\IEEEauthorrefmark{1}Department of Electrical Engineering,
Indian Institute of Technology (IIT) Bombay, Mumbai, MH, 400076, India\\
}%
\thanks{Corresponding author: S. Tallur (e-mail: stallur@ee.iitb.ac.in)
}
}
%
%
%


\IEEEtitleabstractindextext{%
\begin{abstract}
We report the fabrication and characterization results of a simple and low-cost pH sensor fabricated using a graphite pencil to define a working electrode and silver paste to define a reference electrode on scotch tape. The sensor operation is based on potentiometric measurement and thereby insensitive to fabrication variations in shape of the electrode unlike amperometric and chemiresistive measurement techniques. The substrate of the disposable sensor is prepared by pasting scotch tape atop a piece of chart paper, and two types of sensors fabricated with 6B and 2B graphite pencils are tested with three solutions with different pH values. The sensor functions as a passive sensing tag without requiring any external power or stimulus, and the measured sensitivities of the pH sensors fabricated using 2B and 6B pencil carbon electrodes (PCEs) are $-4.54mV/pH$ and $-4.09mV/pH$ respectively. 
\end{abstract}

}


\maketitle

\section{Introduction}\label{AA}

Measuring the acidity (pH) of samples has a wide variety of applications such as environmental monitoring of soil and water samples, quality control and manufacturing in pharmaceutical and chemical industry, manufacturing of food and beverages, healthcare and clinical applications such as blood chemistry etc. \cite{richter2008review,vonau2006ph,khan2017review}. Low cost pH sensitive dyes are good candidates for ubiquitous pH sensing since they also require minimal training or expertise for sensor operation in the field. However, the operator must have a trained eye and perform qualitative comparison of the resulting change in color to a standard color chart, and can thus introduce operator error (repeatability) \cite{b3, lee2006colorimetric}. A glass electrode based potentiometric pH sensor is most commonly used for accurate pH measurements, but is better suited for lab usage as compared to the field due to fragility and packaging costs. Disposable and low cost pH sensors based on emerging materials such as carbon nanotubes (CNTs), graphene, metal oxides etc. \cite{b4,b5,b6,pHreview,fog1984electronic,kaempgen2006transparent,nag2017wearable,jamal2019development} assembled on flexible substrates are better suited for field use. Besides the costs involved in material synthesis, the sensitivity of these sensors is affected by variations in the size of electrodes and weight of solution, purity of the materials, or variations in the manufacturing process.

While the cost of preparation of such materials could be further scaled down, carbon based sensors using pencil carbon electrodes (PCEs) have been reported for a variety of applications \cite{b7,karimi2015novel,rezaei2016modified,hoque2017pencil} and promise an inexpensive alternative for pH sensing \cite{b1,awasthi2018novel}. While a chemiresistive implementation relaxes auxiliary instrumentation overhead for such sensors \cite{b7}, the sensitivity may be affected by sensor packaging which could alter the effective electrode area. This challenge can be overcome by assembling pH sensors on paper substrates and employing a potentiometric technique to perform the measurement. Pencil graphite also contains kaolin clay and oxygen functional groups such as hydroxyl, carboxyl, quinonyl etc. that make the graphite a p-type semiconductor. Ions present in the solution under test interact with the functional groups and change the surface potential of the graphite electrode.  The H$^+$ ions in an acidic solution will accept the electrons in the graphite, leading to an increase in the number of majority carriers and therefore an increase in the potential of the graphite electrode with respect to the electrolyte. In an alkaline solution, the OH$^-$ ions will donate electrons to the graphite and decrease the number of majority carriers, therefore reducing the potential of the graphite electrode with respect to the electrolyte. The sensitivity of the electrode to pH also depends on the binding of the ions to the electrode \cite{b1,b8}. Here we report an improved version of such a scheme, and demonstrate a manually assembled pH sensor with PCE working electrode and silver paste based reference electrode on scotch tape on paper. The paper substrate is used as a mechanical support for the sensor, and the scotch tape protects the paper substrate from absorbing the solution under test, thereby improving the sensor reliability. The potential difference between the two electrodes varies with the pH of the solution surrounding the electrode assembly. The potential difference is measured using a handheld digital multimeter for sensors fabricated using 2B and 6B graphite pencils, that display sensitivities of $-4.54mV/pH$ and $-4.09mV/pH$ respectively.

The remainder of the paper is structured as follows: the methodology followed in this work: sensor fabrication, test solution preparation and experimental procedure are described in section \ref{CC}. Experimental results and observations are presented in section \ref{HH}, followed by future work and conclusion in section \ref{II}.

\section{Methodology}\label{CC}

\subsection{Sensor design and fabrication}\label{DD}

\begin{figure}
    \centering
  \includegraphics[width=\linewidth]{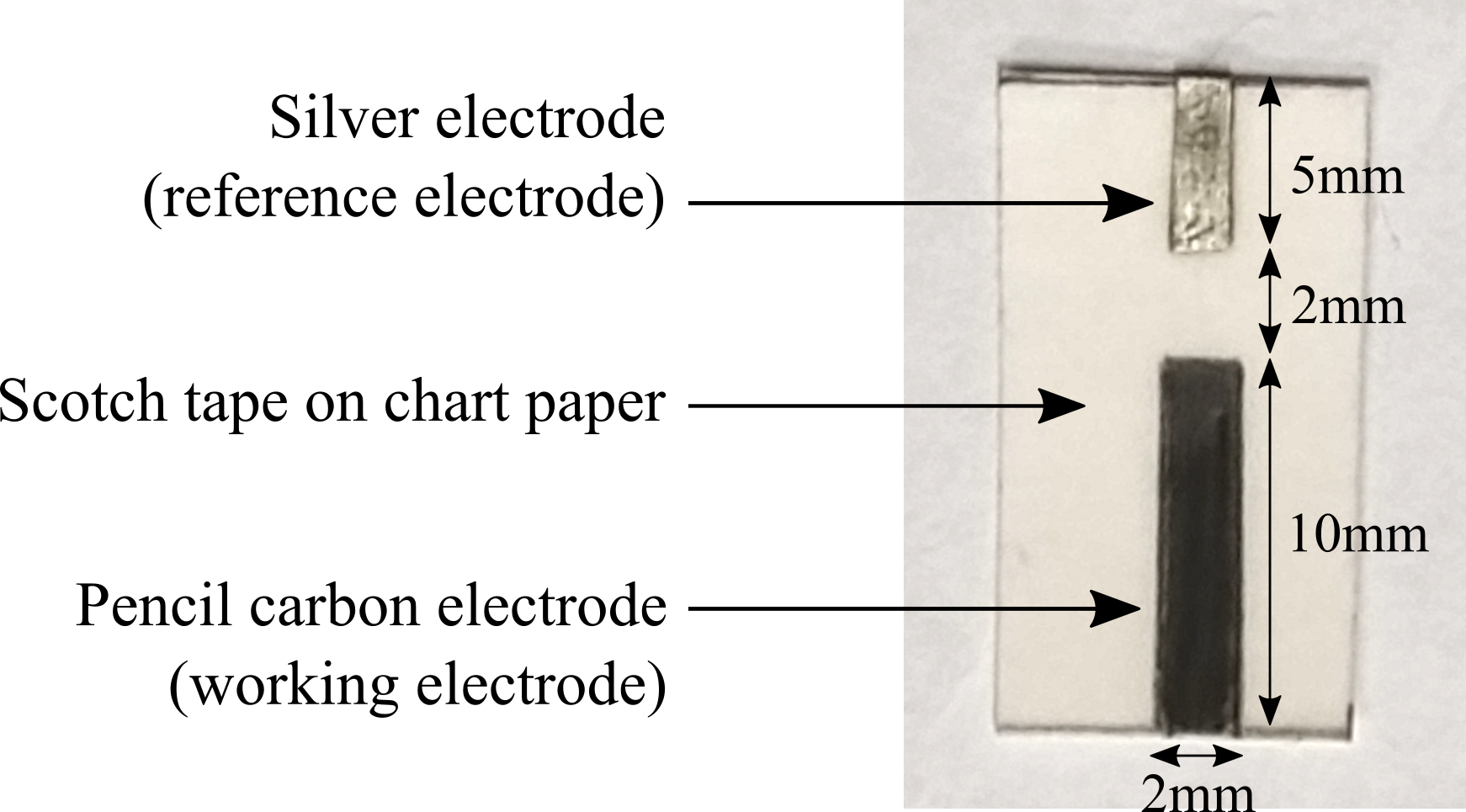}  
    \caption{A photograph of a pH sensor fabricated using a reference electrode prepared using silver paste, and a pencil carbon electrode (PCE) as working electrode. The dimensions of various components of the sensor are highlighted.}
    \label{fig:sensor}
\end{figure}

The substrate of the sensor is prepared by sticking scotch tape on a piece of white colored chart paper of dimension $1.7cm \times 1 cm$. The scotch tape helps preserve the sensor by preventing the paper substrate from absorbing the test solution that is dispensed on top of the sensor, and therefore preventing bending and expansion of the substrate from occurring during the experiment. The reference electrode is prepared using silver paste dispensed on a piece of chart paper of size $5mm \times 2mm$, which is in turn pasted on top of the scotch tape. The working electrode of size $10mm \times 2mm$ is prepared by directly scribbling 2B/6B graphite pencils on the scotch tape, while maintaining a separation of $2mm$ between the two electrodes, where the sample to be tested is dispensed with a micro-pipette. The silver paste is inert and does not react with the solution under test, and helps measure the shift in potential of the working electrode. Using silver paste also makes the electrode relatively more immune to Cl$-$ ions in the solution under test, as compared to conventional Ag/AgCl reference electrodes \cite{b1}. A photograph of an assembled sensor is shown in Figure~\ref{fig:sensor}. While the sensors used in the study reported here were manually assembled, the design and process are compatible with screen printing, laser cutting and batch fabrication manufacturing techniques.

\subsection{Preparation of buffer solutions}\label{FF}

For testing the paper based pH sensors, we prepared three different buffer solutions of pH $4.01$, $7.5$ and $9.0$, to cover a wide range of pH sensing applications \cite{vonau2006ph}. It is essential that the solutions retain their pH over the duration of the experiment and therefore it is desirable to use buffer solutions for conducting these experiments. The volumes of reagents for the solutions are obtained from stoichiometry calculations. We prepared a solution of $155mg$ boric acid (H$_3$BO$_3$) and $0.865ml$ of $1M$ sodium hydroxide (NaOH), to which we add deionized water (DI water) to prepare $25ml$ buffer solution of pH $9.0$. Similarly we prepare $25ml$ of pH $7.5$ buffer solution using $5.63g$ disodium hydrogen phosphate (Na$_2$HPO$_4$), $3.25ml$ hydrochloric acid (HCl) and DI water. The buffer solution of pH $4.01$ used for characterizing sensor response to acidic solutions is $20ml$ of a standard buffer solution of potassium hydrogen phthalate, commonly known as KHP (C$_8$H$_5$KO$_4$), DI water and red dye. The pH values of the buffer solutions are validated independently using an Oakton pH150 portable pH meter (Cole-Parmer).


\subsection{Experimental procedure}\label{GG}

\begin{figure}
    \centering
  \includegraphics[width=\linewidth]{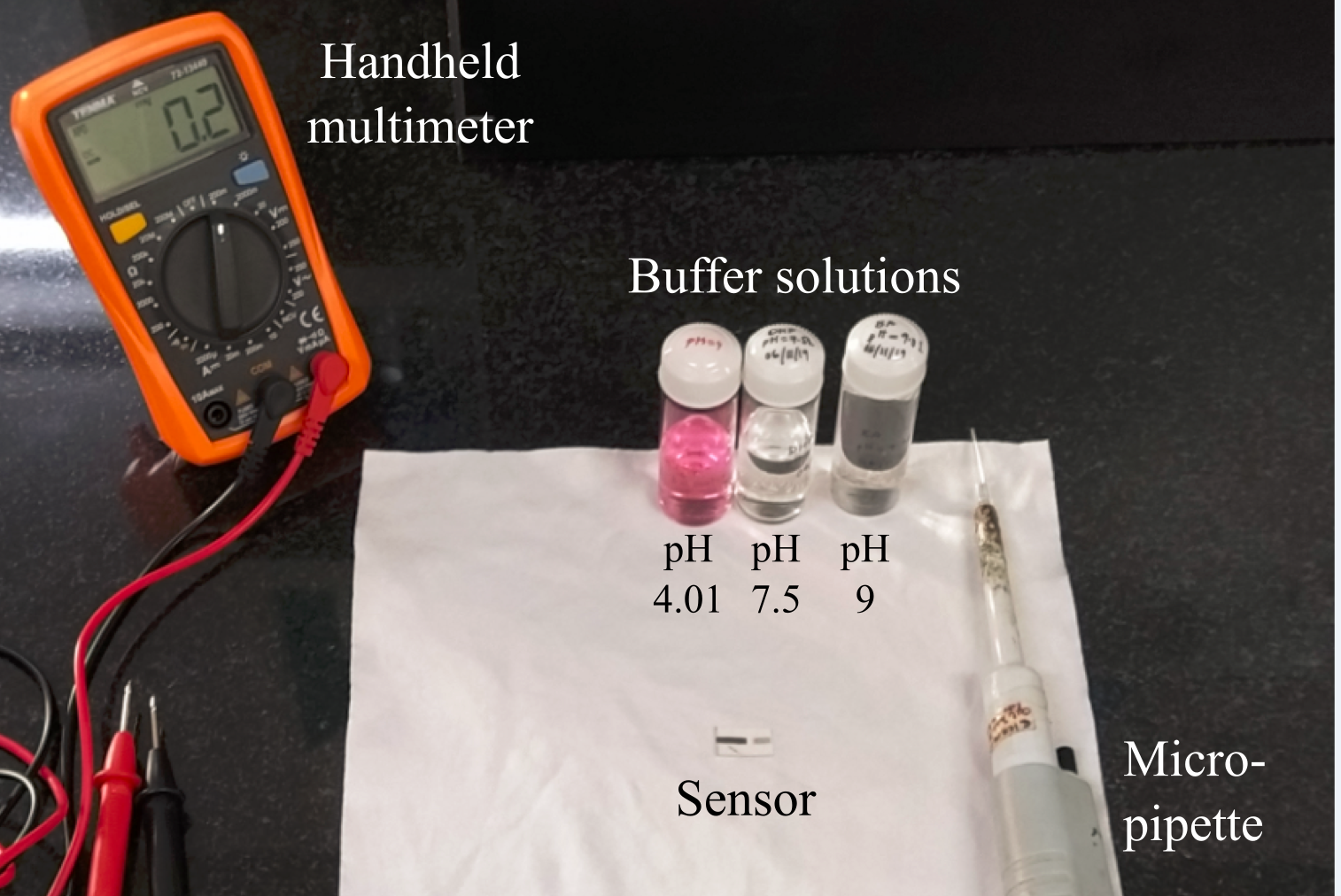}  
    \caption{Photograph of the experimental setup showing an assembled sensor, handheld multimeter for measuring the potential difference between working electrode and reference electrode, buffer solution vials and micro-pipette used for dispensing the solutions on to the pH sensors.}
    \label{fig:setup}
\end{figure}

Figure \ref{fig:setup} shows a photograph of the work table prior to conducting the experiment. The potential difference between reference electrode and working electrode is measured with a handheld digital multimeter. When there is no test solution dispensed on the sensor, the multimeter measures the offset voltage of the sensor. We measured an average offset voltage of $-0.26mV$ for the sensors used in this study. To measure the response of the sensor to a buffer solution, $10\mu l$ of the buffer solution is dispensed between the two electrodes of a solution using a micro-pipette. The potential difference is measured $300 s$ after dispensing the sample, in order to allow the system to reach equilibrium. The offset voltage is subtracted from the potential difference measured when the solution is present on the sensor. Considering that the motivation of this work is the development of low cost, disposable pH sensors, we prepared 15 sensors each using 2B and 6B pencils, and use 5 sensors of each type for measuring the pH of every buffer solution. This experiment method eliminates errors due to improper cleaning and residues if every sensor were to be reused and measured with all three buffer solutions. However since each individual sensor is only used once, the sensitivity measurements reported in the next section are prone to unit-to-unit variations for measurements conducted on various units of a particular type of sensor. 

\section{Results and Discussion}\label{HH}
\begin{figure}
    \centering
  \includegraphics[width=\linewidth]{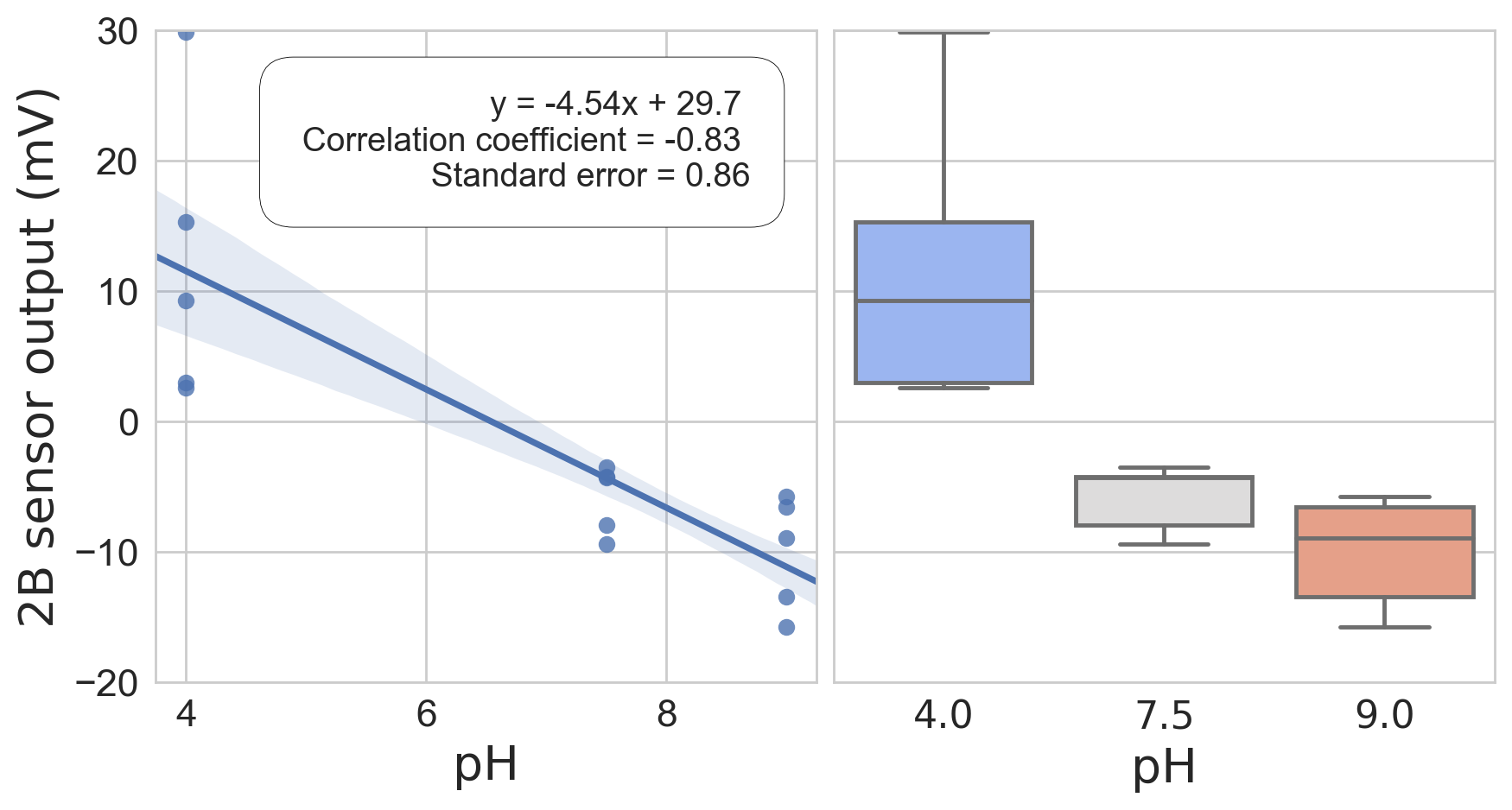}  
    \caption{Distribution of measured sensor output for pH sensor prepared using 2B pencil for test solutions with pH = 4.01, 7.5 and 9, shown as a linear regression plot (left) and box-plots (right). The sensitivity obtained from linear regression is $-4.54mV/pH$.}
    \label{fig:2B}
\end{figure}

\begin{figure}
    \centering
  \includegraphics[width=\linewidth]{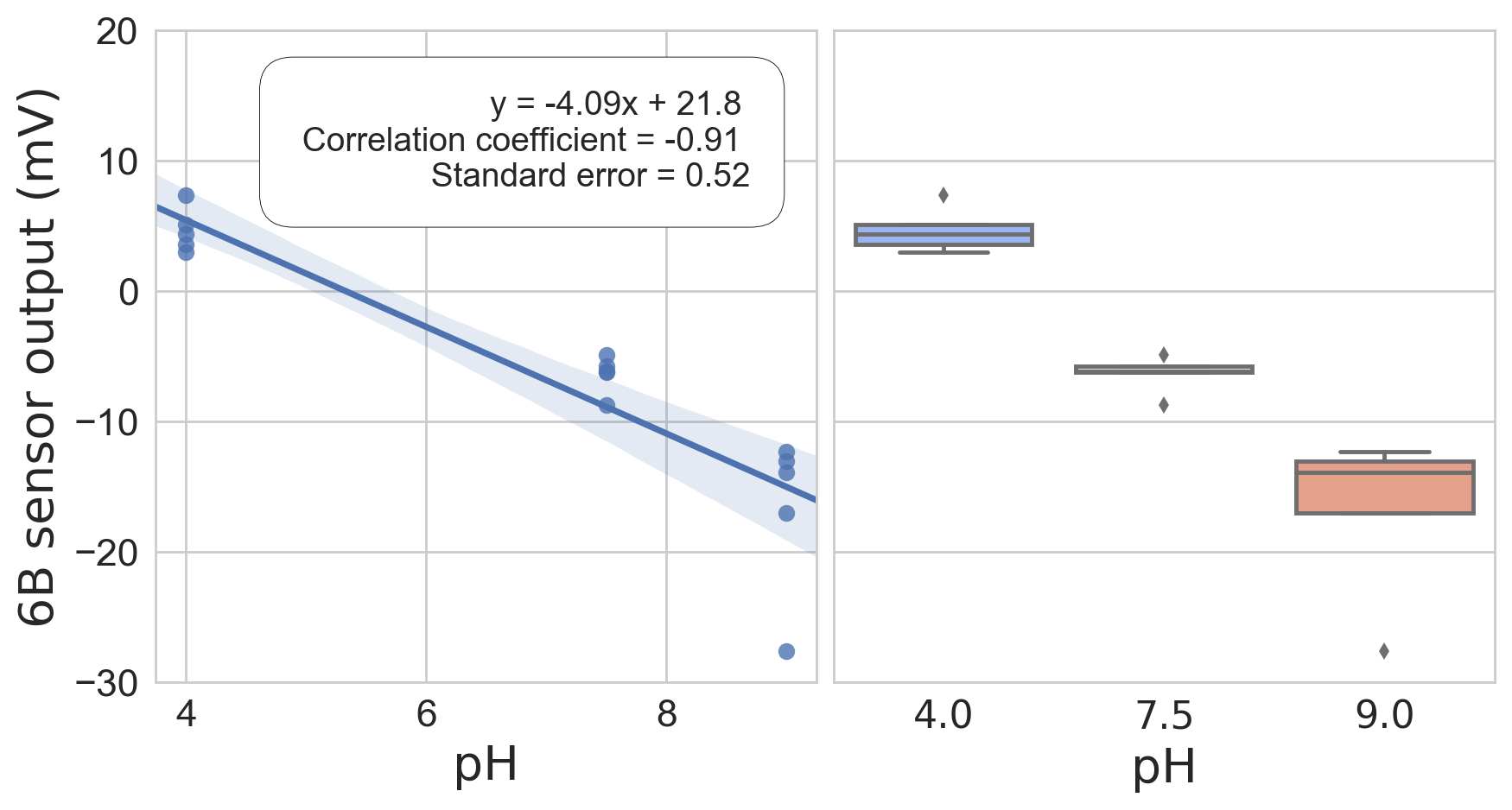}  
    \caption{Distribution of measured sensor output for pH sensor prepared
    using 6B pencil for test solutions with pH = 4.01, 7.5 and 9, shown as a linear regression plot (left) and box-plots (right). The sensitivity obtained from linear regression is $-4.09mV/pH$. The sensors prepared with 6B pencil show lesser standard error in measurements $(0.52)$ and better correlation coefficient in the linear regression plot $(-0.91)$ as compared to sensors prepared using 2B pencil $(0.86$ and $-0.83$ respectively$)$.}
    \label{fig:6B}
\end{figure}

The measured sensor outputs for sensors fabricated using 2B and 6B pencils are shown in Figures~\ref{fig:2B} and \ref{fig:6B} respectively. Since every sensor was used only once, we perform linear regression (using seaborn library in Python) to estimate the sensitivities of both types of sensors. The confidence intervals shown in the translucent band in the regression plots denotes a confidence interval of $68\%$, corresponding to the standard deviation. The sensors prepared using 6B pencil show lower magnitude of sensitivity $(-4.09mV/pH)$ as compared to the sensors fabricated using 2B pencil $(-4.54mV/pH)$. However, the standard error (spread) is lower for sensors prepared using 6B pencil $(0.52)$ and the regression correlation coefficient is also better $(-0.91)$ as compared to sensors prepared using 2B pencil $(0.86$ and $-0.83$ respectively$)$ . This may be attributed to the higher clay content in 2B pencils $(20\%)$ as compared to 6B pencils $(10\%)$ \cite{sousa2000observational}. Graphite pencil lead containing a larger amount of clay has higher resistivity because clay is an insulator. The higher resistivity could interfere with the measurement of the potential change at the electrode surface in contact with the solution. Due to the disposable nature of the experiment, as well as manual fabrication process, the measurements suffer from unit-to-unit variations. Additionally, it is worth noting that the graphite electrode is formed by weak Van der Waals force holding together a network of graphite that is formed by scribbling the pencil on scotch tape. The graphite network (and hence the surface condition) at the PCE changes upon dispensing a sample, and without any additional binding agent, these variations introduce additional spread in the measurements.

\section{Conclusion and future work}\label{II}

We present a low-cost method to fabricate PCE based pH sensors on a flexible paper substrate. The paper substrate is protected from the solutions under test using scotch tape. The sensor is used in a potentiometric measurement scheme, that is better suited for flexible substrates as compared to amperometric techniques. The sensor requires no additional circuitry or external power or stimulus, and can therefore be integrated as a passive pH sensing tag into any instrumentation. We measure sensitivities of $-4.54mV/pH$ and $-4.09mV/pH$ for sensors prepared using 2B and 6B graphite pencils respectively.

The measurements reported here are limited by resources available to the authors for conducting the experiments. The measurements have inherent unit-to-unit variations due to manual fabrication process, surface changes at the working electrode due to weak binding of the graphite network on the scotch tape, and disposable nature of the experimental methodology. The sensitivty of a population of the sensors in presence of all these variations is estimated through linear regression, wherein the sensors prepared using 6B pencil show better correlation and measurement spread. This is attributed to lower clay content in the 6B pencil that results in reduced surface resistivity of the graphite, and therefore lesser contact resistance while making measurements. Future work may focus on improving the sensor fabrication process using combination of laser cutting the substrate and screen printing electrodes, imaging and characterizing the microstructure of the pencil lead and analyzing its graphite and clay content through spectroscopy to identify correlations to the sensor performance, investigating appropriate binding techniques to improve adhesion of the graphite in the PCE to the substrate, as well as performing additional measurements with buffer solutions with various pH values.


\section*{Acknowledgement}

The authors thank Prof. Arindam Sarkar at the Department of Chemical Engineering, IIT Bombay for access to laboratory facilities for pH sensor characterization; Mr. M. Shariq Anwar, Ph.D. student in the Department of Chemical Engineering, IIT Bombay for assistance in preparation of buffer solutions; and Ms. Ruchira Nandeshwar,  Ph.D. student in the Department of Electrical Engineering, IIT Bombay for assistance with performing experiments in the Biosensors Lab at IIT Bombay. The components and consumables required for the project were provided by the Wadhwani Electronics Lab (WEL) at IIT Bombay.

\newpage
\bibliographystyle{IEEEtran}
\bibliography{ref}

\end{document}